\documentclass[12pt]{article}
\usepackage[utf8]{inputenc}
\usepackage{amsfonts}
\usepackage{enumitem}
\usepackage{graphicx}
\usepackage{cite}
\usepackage{color}
\usepackage{amsmath,amssymb,amsthm}
\usepackage{mathrsfs}
\usepackage{slashed}
\usepackage[english]{babel}
\usepackage[left=1cm,top=1cm,bottom=1cm,right=1cm,nohead,nofoot]{geometry}

\def \be {\begin{equation}}
\def \ee {\end{equation}}
\def \bea {\begin{eqnarray}}
\def \eea {\end{eqnarray}}
\def \nn {\nonumber}

\def \rr {\raise.35ex\hbox{\small $\prime$}\kern-.17em{\mbox{\large $\imath$}}}

\def \dels {\partial\kern-.6em /\kern.1em}
\def \As {{A\kern-.5em / \kern.5em}}
\def \Ds {D\kern-.7em / \kern.5em}

\def \ks {k\kern-.5em /}
\def \ls {l\kern-.5em /}

\newcommand{\ci}[1]{}
\newcommand{\ba}{\begin{eqnarray}}
\newcommand{\ea}{\end{eqnarray}}
\newcommand{\bal}{\begin{align}}
\newcommand{\eal}{\end{align}}
\newcommand{\bay}[1]{\left(\begin{array}{#1}}
\newcommand{\eay}{\end{array}\right)}

\reversemarginpar
\newcommand{\hide}[1]{}

\makeatletter

\newlist{axioms}{enumerate}{2}
\setlist[axioms,1]{label=\textbf{A\arabic{axiomsi}.}, ref=A\arabic{axiomsi}}
\setlist[axioms,2]{label=\textbf{A\arabic{axiomsi}\rlap{\myEnumCounter{axiomsii}}.},%
                   ref=A\arabic{axiomsi}\myEnumCounter{axiomsii},%
                   align=parleft,%
                   leftmargin=0em,%
                   itemsep=1.4ex,%
                   before={\stepcounter{axiomsi}}}

\usepackage{tikz,pgf}
\usetikzlibrary{shapes}
\usetikzlibrary{calc}
\usetikzlibrary{decorations.pathmorphing}
\usetikzlibrary{decorations.pathreplacing,shapes.misc}
\usetikzlibrary{positioning}
\usetikzlibrary{arrows}
\usetikzlibrary{decorations.markings}

\tikzset{snake it/.style={decorate,decoration={snake,segment length=1.5mm, amplitude=.3mm}}}
\tikzset{biggerarrow/.style={
    decoration={markings,mark=at position 1 with {\arrow[scale=1.5]{>}}},
    postaction={decorate},
    shorten >=0.4pt}}
\tikzset{arrow at middle/.style={decoration={
    markings,
    mark=at position 0.5 with {\arrow{>}}}}}

\begin{document}

\begin{titlepage}

\begin{center}

\hfill
\vskip .2in

\textbf{\LARGE
Supergravity with Doubled Spacetime Structure
\vskip.3cm
}

\vskip .5in
{\large
Chen-Te Ma$^a$ \footnote{e-mail address: yefgst@gmail.com} and Franco Pezzella$^b$ \footnote{e-mail address: franco.pezzella@na.infn.it}
\\
\vskip 1mm
}
{\sl
${}^a$
Department of Physics and Center for Theoretical Sciences, \\
National Taiwan University,\\ 
Taipei 10617, Taiwan, R.O.C.\\
$^b$ 
Istituto Nazionale di Fisica Nucleare - Sezione di Napoli,\\
Complesso Universitario di Monte S. Angelo ed. 6,\\
 via Cintia,  80126 Napoli, Italy.
}\\
\vskip 1mm
\vspace{40pt}
\end{center}
\begin{abstract}
Double Field Theory (DFT) is a low-energy effective theory of a manifestly $O(D,D)$ invariant formulation of the closed string theory when toroidally compactified dimensions are present. The theory is based on a doubled spacetime structure and, in order to preserve the gauge symmetry provided by the invariance under generalized diffeomorphisms, a constraint has to be imposed on fields and gauge parameters. In this paper, we propose a DFT-inspired Supergravity by using a suitable {\em star product} that implements such constraint and the corresponding algebraic structure is explored. We get a consistent DFT in which also an orthogonality condition of momenta is necessary for having a closed gauge algebra.
In constructing this theory, we start from the simplest case of doubling one spatial dimension where the action is uniquely determined, without ambiguities, by the gauge symmetry. Then, the extension to the generic $O(D, D)$ case is presented. The result is consistent with the closed string field theory. 
\end{abstract}

\end{titlepage}

\section{Introduction}
\label{1}
The $O(d,d;{\mathbb Z})$ T-duality is an exact symmetry of a closed string wrapped around the $d$ non-contractible cycles of a $d$-torus $T^{d}$ in a $D$-dimensional spacetime  ${\mathbb R}^{n-1,1} \times T^{d}$ where ${\mathbb R}^{n-1,1}$ is the $n$-dimensional Minkowski space  ($D=n+d$). The duality makes the string physics at a very small scale indistinguishable from the one at a large scale, and also gives an evidence that ordinary geometry could possibly be broken down at the string scale.  Double Field Theory (DFT) \cite{Hull:2009mi, Siegel:1993xq}  is a field theoretical approach that incorporates such symmetry.  It develops, therefore, a stringy geometry, which makes T-duality manifest. Its fields, i.e. the gravity, antisymmetric tensor and dilaton fields are defined in ${\mathbb R}^{n-1,1} \times T^{2d}$ where the {\em double torus} $T^{2d}$ contains the original space-time torus $T^{d}$ with its ordinary coordinates $x^{i}$ ($i=1, 2, \cdots, d$) and the dual torus, related to $T^{d}$  by the T-duality and parameterized by the dual coordinates $\tilde{x}_{i}$ associated with winding excitations. Actually, the DFT formalism is based on a doubling of all the $D$ coordinates of the $D$-dimensional space ${\mathbb R}^{n-1,1} \times T^{d}$ with the full duality group $O(D,D)$ that is broken to the subgroup $O(d,d; {\mathbb Z})$ preserving the periodic boundary conditions along the $d$ compact dimensions of $T^{d}$ together with the condition of no winding in the non-compact directions.

DFT is invariant under generalized diffeomorphisms if the so-called {\em weak constraint}, i.e. $\Delta f \equiv    \partial_{m} \tilde{\partial}^{m}  f =0  \,\,\, (m=1, 2, \cdots, D)$, is satisfied by the fields and gauge parameters. Consequently, one can introduce a projector in order to restrict  an arbitrary field or a gauge parameter to the kernel of the differential operator $\Delta$. This projection can be promptly embodied by the suitable definition of a {\em star product operator}. Upon introducing such a projector, one can construct an action defined in terms of restricted fields. Actually, the use in the action of restricted fields seems to avoid the {\em strong constraints}. The latter are the extension of the weak constraint to the product of fields, introduced for having a consistent manifestly background independent DFT action \cite{Hohm:2010jy, Hohm:2015ugy} with the equivalent generalized metric formulation \cite{Hohm:2010pp} and, furthermore, to make DFT supersymmetric \cite{Hohm:2011nu}. 
The star product that is going to be defined here is not associative. In general, the non-associativity property may imply non-closure of the gauge algebra. But we can show that it is possible to get a closed gauge algebra under particular conditions that will be discussed later.

In this paper, we exhibit the properties of a supergravity in the doubled spacetime but with its fields restricted in the kernel of $\Delta$  through the star product. The starting point is the action in ref. \cite{Hohm:2010pp}, but it is important to stress that the use of restricted fields avoids the strong constraints that otherwise would be necessary for having gauge invariance. We first consider the $d=1$ case and explore the algebraic structure of the gauge symmetry. The result is that the involved fields are required to have momenta orthogonal to each other. This guarantees the closure of the gauge algebra. In this theory, the action can therefore be uniquely determined by the gauge symmetry and results to be the sum of the ordinary supergravity and the dual supergravity. We also find that the action at the cubic order is exactly consistent with the closed string field theory if a non-trivial boundary is absent. 

The extension to the $O(D, D)$ case is made also by assuming that all momenta are orthogonal to each other, implying also in this case the closure of the gauge algebra. This assumption is supported by the agreement, at the cubic order, with the action obtained from the closed string field theory. The role played by the orthogonality of momenta, that can be derived from the triple product, makes our considerations and motivations different from the ones in \cite{Lee:2015qza}.  

We first discuss the projector  in Sec.~\ref{2}. The properties of the new supergravity action are illustrated in Sec.~\ref{3} and \ref{4}. Finally, in Sec.~\ref{5} we report the conclusions and our outlook. Further technical details will be reported in a forthcoming paper \cite{MP}. 

\section{Properties of the Projector}
\label{2}
In order to impose the weak constraint in DFT, a projector \cite{Hull:2009mi} is necessary to force fields and gauge parameters to live in the kernel of the operator  $\Delta \equiv  \partial{_m} \tilde{\partial}^{m} $ with $m=1, 2, \cdots, D$. This requirement, in turn, is necessary for preserving the invariance under generalized diffeomorphisms, which constitutes the gauge symmetry.

 The weak constraint on arbitrary fields and gauge parameters is equivalently written in an $O(D,D)$ covariant form as follows:
\bea
\partial_M\partial^M f=0,   \,\,\,\,\,\,\,\,\,\,\,\,\, \mbox{with}\,\,\,\, M=1, 2, \cdots, 2D,
\eea
where 
\bea
\partial_M\equiv 
 \begin{pmatrix} \,\tilde{\partial}^m \, \\[0.6ex] {\partial_m } \end{pmatrix}  \,\,\,\,\,\,\,\,\,\,\,\,\, \mbox{with}\,\,\,\, m=1, 2, \cdots, D.
\eea 
Here, we denote the $O(D, D)$ or doubled indices through the use of capital letters and the non-doubled indices through the use of lower-case letters. The doubled indices are raised or lowered by the constant $O(D,D)$ invariant metric
\bea
\eta_{MN}= \begin{pmatrix} 0& I \\ I& 0 \end{pmatrix}.
\eea
%In this notation one can consider a theory with both compact and non-compact dimensions.
 For a generic double field $A$, one can introduce a Fourier series along the dimensions of the doubled spacetime as follows:
\bea
A\equiv\sum_K A_K e^{iKX},
\eea
where $KX\equiv K^MX_M$. Here we define: $X^{M} \equiv (\tilde{x}_{m}, x^{m})$ and $K^{M} \equiv (p_{m}, w^{m})$, being $p_{m}$ the momentum along the $m$-th dimension and $w^{m}$ the corresponding winding number ($m=1, 2, \cdots,D$).  %Non-doubled dimensions are denoted by Greek indices. 

We then introduce the star product to embody the projection on the kernel of $\Delta$ or the weak constraint in the following way:
\bea
A*1\equiv\sum_K A_K e^{iKX}\delta_{KK,0}     \label{proj}\ ,
\eea
where $KK\equiv K^MK_M$. In other words, the star product imposes a weak constraint in $A$, since it is straightforward to show that:
\[
\partial_M\partial^M (A*1)=0.
\]
One can also define the star product of weakly constrained fields $A$ and $B$  as follows:
\bea
A*B=B*A=\sum_{K, K^{\prime}}A_KB_{K^{\prime}} e^{i (K+K^{\prime} ) X} \delta_{KK, 0} \, \delta_{K^{\prime}K^{\prime}, 0} \, \delta_{KK^{\prime}, 0}.
\eea
 Generically, the star product does not exhibit associativity, since:
 \begin{eqnarray}
 (A*B)*C  \neq A*(B*C)
 \end{eqnarray}
  but one can have associativity under integration as follows:
\bea
\int dX\ A*(B*C)=\int dX\ (A*B)*C.
\eea
The triple product also satisfies the weak constraint:
\bea
\partial^M\partial_M\big(A*(B*C)\big)=\partial^M\partial_M\big((A*B)*C\big)=0
\eea
and this property can be extended to the product of an arbitrary number of fields and gauge parameters. 

%Of course, since DFT is  based on a manifest $O(D, D)$ invariance, being the theory formulated on a torus which has periodic boundary conditions, the projector defined in  eq. (\ref{proj}) can be used. In order to obtain the local physics of the non-compact double space, we can consider an infinite size of the compact torus as in lattice theory.

\section{ $d=1$}
\label{3}
In the case of  $d=1$, the constraints 
\bea
KK=0 \qquad K^{\prime}K^{\prime}=0 \qquad (K+K^{\prime})K^{\prime\prime}=0
\eea
can be solved as follows:
\bea
K=aK^{\prime}=bK^{\prime\prime},
\eea
where $a$ and $b$ are arbitrary constants. It naturally follows that
\bea
K^{\prime\prime}K^{\prime\prime}=0.
\eea
All the previous relations make all the momenta orthogonal to each other and, consequently, make star product associative. Furthermore, arbitrary functions along the compact dimension, in this case, have the following form:  
\vspace{.2cm}
\bea
f=\sum_K f_K e^{iKX}=\sum_{p}f_{p}  e^{ip^{m}x_{m}}+\sum_{\omega}f_{\omega} e^{i\omega^m\tilde{x}_m}. \label{fatt}
%f=\sum_K f_K e^{iKX}=\sum_{p^{\mu}, p_i}f_{p^{\mu}, p_i}  e^{ik^{\mu}x_{\mu}+ip_ix^i}+\sum_{p^{\mu},\omega^i}f_{p^{\mu},\omega^i} e^{ip^{\mu}x_{\mu}+i\omega^i\tilde{x}_i}. \label{fatt}
%\nn\\
\eea
\vspace{.1cm}

Therefore, all fields and gauge parameters become the sum of the contribution coming from the ordinary coordinates and the contribution coming from the dual coordinates. 
 %The action in the case of the $O(1, 1)$  DFT can be determined without any ambiguities. 

More explicitly, DFT for $d=1$ is defined by using the star product in the following action \cite{Hohm:2010pp} formulated in terms of the generalized metric ${\cal H}$: \bea
\label{acg}
S_{DFT} &=& \int dx \ d\tilde x  \
   e^{-2d}*\Big(\frac{1}{8}{\cal H}^{MN}*\partial_{M}{\cal H}^{KL}*
  \partial_{N}{\cal H}_{KL}-\frac{1}{2}
  \,{\cal H}^{MN}*\partial_{N}{\cal H}^{KL}*\partial_{L}
  {\cal H}_{MK}
\nn\\
  &&-2\partial_{M}d*\partial_{N}{\cal H}^{MN}+4{\cal H}^{MN}*\,\partial_{M}d
  *\partial_{N}d \Big),   \label{SDFT}
 \eea
 where
 \bea
{\cal H} \ \equiv \
  \begin{pmatrix}    g-B*g^{-1}*B & B*g^{-1}\\[0.5ex]
  -g^{-1}*B & g^{-1}\end{pmatrix}, \qquad
  e^{-2d}\equiv\sqrt{-g}*e^{-2\phi},
  \eea
and $g$ is the metric and $d$ is the dilaton both defined on the doubled spacetime. The gauge transformations are:
\bea
\delta_{\xi} d&=&-\frac{1}{2}\partial_M\xi^M+\xi^M*\partial_M d,
\nn\\
\delta_{\xi} {\cal H}^{MN}&=&\xi^P*\partial_P{\cal H}^{MN}+(\partial^M\xi_P-\partial_P\xi^M)*{\cal H}^{PN}+(\partial^N\xi_P-\partial_P\xi^N)*{\cal H}^{MP},
\nn\\
\eea
and they exhibit the closure property as
\bea
[\delta_{\xi_1}, \delta_{\xi_2}]=-\delta_{[\xi_1, \xi_2]_C},
\eea
where the $C$-bracket is defined by
\bea
[\xi_1, \xi_2]_C^M=\xi_1^N*\partial_N\xi_2^M-\xi_2^N*\partial_N\xi_1^M-\frac{1}{2}\eta^{MN}\eta_{PQ}\xi_1^P*\partial_N\xi_2^Q+\frac{1}{2}\eta^{MN}\eta_{PQ}\xi_2^P*\partial_N\xi_1^Q.
\nn\\
\eea

Due to the the separation in two parts each of them respectively dependent on $x$ and $\tilde{x}$, shown in eq. (\ref{fatt}) and induced by the star product, %
%actually generates that kinds of splitting for fields and gauge parameters in DFT. Therefore, as already observed,
 the density Lagrangian in the action $S_{DFT}$ can be rewritten as:
\bea
{\cal L}[x, \tilde{x}] = {\cal L}_{1}[x] + {\cal L}_{2}[\tilde{x}]
\eea
where ${\cal L}_{1}$ and ${\cal L}_{2}$ are formally identical to ${\cal L}[x, \tilde{x}]$ in the DFT action (\ref{SDFT}), but with a dependence of the fields respectively on $x$ and $\tilde{x}$ instead of $[x, \tilde{x}]$. 

The star product in the action can actually be replaced by the ordinary product because of the identity
\bea
&&\int dX\ A\cdot B\cdot C\cdots=\int dX\ A*B*C\cdots
\eea 
without considering the eventual non-trivial boundary term. If the non-trivial boundary term appears in DFT, then the Fourier analysis should fail. This situation possibly appears in the non-compact space when we consider a double torus with an infinite size. It is interesting to assume this result as a starting point for shedding light on the meaning of DFT.

\section{ $O(D, D)$ Double Field Theory}
\label{4}
 In this section we extend our discussion to the generic $O(D, D)$ case. In order to find the similar property already discussed in the $d=1$ case, we {\em impose additional conditions} to redefine the star product:
\bea
KK=0, \qquad K^{\prime}K^{\prime}=0, \qquad K^{\prime\prime}K^{\prime\prime}=0,\ \cdots\cdots, KK^{\prime}=0, \qquad KK^{\prime\prime}=0,\ \cdots\cdots.
\eea
and, in particular, this choice implies, also in this case, the associativity property. Recovering this crucial property enforces us to modify the DFT action through the insertion of the star product in order to find the gauge invariant action. Since all momenta are orthogonal to each other, we can ensure the equivalence between the ordinary product and star product to all orders by reducing fluctuations of fields. We also find non-trivial evidence to support our approach. The evidence comes from the orthogonality condition implied by the triple product of fields. Hence, this shows that the action obtained from the closed string field theory can imply the orthogonality condition. At the cubic order, the orthogonality condition does not influence the fluctuations of the fields.
 
The generic $O(D, D)$ DFT is still the one in eq. (\ref{SDFT}) rewritten in terms of the generalized metric, and the gauge transformations are the same as in the $d=1$ case with the same property of closure.

\section{Conclusion and Outlook}
\label{5}
We propose a supergravity in the doubled spacetime by embodying the weak constraint through a suitable projector acting on fields and gauge parameters. This method could also allow us to avoid the no-go of the maximum number of spacetime dimensions when supersymmetric extension is considered. 

We have first considered the DFT for $d=1$.  In this case, the theory is uniquely determined by the gauge symmetry. This has resulted in a well-defined example without ambiguities. The interesting aspect in this case is that the momenta of the fields result to be orthogonal to each other and it implies the closure of the gauge algebra. Hence, the action is not necessarily modified to all orders.

The extension of the algebra to the $O(D, D)$ case has been studied by choosing all momenta, in the Fourier expansion of fields, orthogonal to each other just by analogy with what happens in the $d=1$ case. The non-trivial evidence supporting this orthogonality comes from the orthogonality that holds under integration in the case of cubic action \cite{Hull:2009mi}. In this way, we still have the associativity property of the star product and it implies the closure of the gauge algebra. The DFT with restricted fields can be shown to exhibit the properties of the original double field theory like supersymmetry \cite{Hohm:2011nu} and background independence to all orders \cite{Hohm:2015ugy}. The suitable double sigma model is also easier to be defined by using the star product. This is motivated by the fact that the ordinary product in the expression of the one-loop $\beta$ function should be itself substituted by the star product \cite{Copland:2011wx}, and one can also observe the equivalence of the equations of motion with the off-shell self-duality relation in the $d=1$ case. 

 T-duality in DFT seems to be modified as well by the star product. In fact, such a modification only appears when a non-trivial boundary appears in our theory. This implies that the definition of the non-geometric flux requires some attention in regards the boundary term or the global geometry. Now we can understand why non-single valued fields must be removed to define a consistent action with the non-geometric flux, after performing T-duality \cite{Andriot:2011uh}. T-duality is defined in the non-commuting space so it is also interesting to understand it from the generalized metric defined with the star product \cite{Kamani:2001yd, Jurco:2013upa}.

One subtle question is why we need to choose all momenta orthogonal to each other. The relation is motivated from the $d=1$ case, and we find that the condition can appear in higher dimensions from the triple product. We are still interested in choosing this orthogonality relation because, in this way, we can find all the necessary physical conditions required in a supergravity theory. Thus, we wonder if we may have an interesting connection between the world-sheet space and target space similar to the connection provided by the conformal symmetry in the ordinary string theory. If this condition is general in DFT, then one could see it also from closed string field theory beyond the cubic order. Therefore, imposing the orthogonality can be seen as finding a loophole to give different interpretations to the results of the closed string field theory without violating any widely accepted results.

\section*{Acknowledgments}
We would like to thank Martin Cederwall, Olaf Hohm, Kanghoon Lee, Jeong-Hyuck Park and Masaki Shigemori for interesting discussions and useful suggestion. Especially, Chen-Te Ma would like to thank Nan-Peng Ma for his suggestion and encouragement. 

Moreover, both of the authors would like to acknowledge the hospitality of the APCTP  in Pohang (Korea)  during the workshop {\em Duality and Novel Geometry in M-Theory}. F.P. also thanks the Simons Center and the Galileo Galilei Institute for hosting him for the Summer Workshop in Mathematics and Physics and the {\em Supergravity: what next?} Workshop, respectively.

%\appendix

\end{document}